\documentclass{tlp}
\usepackage{amsmath, amssymb}
\usepackage{latexsym}

\newtheorem{definition}{Definition} % [section]
\newtheorem{example}{Example} % [section]
\newtheorem{theorem}{Theorem} % [section]
\newtheorem{lemma}{Lemma} % [section]
\newtheorem{corollary}{Corollary} % [section]
\newcommand{\logic}[1]{\ensuremath{\mathbf{#1}}}
\newcommand{\calif}[1]{\ensuremath{\mathcal{#1}}}
\newcommand{\set}[1]{\left\{#1\right\}}
\newcommand{\lthen}{\rightarrow}
\newcommand{\liff}{\leftrightarrow}
\newcommand{\Gi}[1][\mathnormal{i}]{\ensuremath{\logic{G}_{#1}}}
\newcommand{\lif}{\leftarrow}
\newcommand{\sig}{\mathcal{L}}
\newcommand{\lxnot}{{\sim}}
\newcommand{\opa}{\begin{math} \oplus_{1} $ $  \end{math}}

\submitted{14 February 2005}
\revised{28 October 2005, 11 August 2006}
\accepted{20 September 2006}

\begin{document}

\title[Updates in Answer Set Programming]{Updates in Answer Set Programming: \\ An Approach Based on Basic Structural Properties}
\author[M. Osorio et al.]{MAURICIO OSORIO and V\'ICTOR CUEVAS \\
Universidad de las Am\'ericas Puebla. \\
Sta. Catarina M\'artir, Cholula, Puebla \\
72820 M\'exico\\
\email{\{osoriomauri, victorcuevasv\}@gmail.com}\\
}

\maketitle

\begin{abstract}
We have studied the update operator \opa defined for update sequences by Eiter \textit{et al.} without
tautologies and we have observed that it satisfies an interesting property\footnote{This paper
can be considered as an extended and revised version of the papers presented in \cite{oz03,zac04}.}.
This property, which we call Weak Independence of Syntax (WIS), is similar to one of the postulates proposed by 
Alchourr{\'o}n, G{\"a}rdenfors, and Makinson (AGM); only that in this case
it applies to nonmonotonic logic.
In addition, we consider other five additional basic properties about update programs
and we show that \opa satisfies them.
This work continues the analysis of the AGM postulates with respect to
the \opa operator under a refined view that considers $\logic{N_2}$ as a monotonic
logic which allows us to expand our understanding of answer sets. Moreover,
$\logic{N_2}$ helped us to derive an alternative definition of \opa avoiding the use
of unnecessary extra atoms.\\

\end{abstract}

\section{Introduction}\label{introduction}

In recent years, a lot of theoretical work has been devoted to explore the
relationships
between intuitionistic logic and answer set programming (ASP) \cite{LiPeVa01,lopstr01,tplp04,apal04}. These
results have provided a characterization of ASP by intuitionistic logic as
follows: a literal is entailed by a program in the answer set semantics
if and only if it belongs to every intuitionistically complete and
consistent extension of the program formed by adding only negated literals \cite{tplp04}.
The idea of these completions using intermediate logics in general is
due to Pearce \cite{Pea99}.
This logical approach provides the foundations to define the notion of nonmonotonic
inference of any propositional theory (using the standard connectives)
in terms of a monotonic logic (namely intuitionistic logic), see
\cite{LiPeVa01,lopstr01,tplp04}.
As Pearce et al. noticed, if we include strong negation we just have to
move to Nelson logics \cite{LiPeVa01}. We select here $\logic{N_2}$ logic,
because it characterizes adequately strong equivalence.
A formalization of $\logic{N_2}$ can be obtained from intuitionistic
logic by adding the axiom schema $F \vee (F \rightarrow G) \vee \neg G$
and the axioms of Nelson logic.

When an intelligent agent obtains new knowledge consistent with its
current information that knowledge is added to the agent's knowledge base. If however
the new knowledge contradicts the agent's current information, a method for incorporating
this new knowledge must be developed. In this paper we present our approach to update nonmonotonic knowledge bases
represented as logic programs under the answer set semantics \cite{GelLif88,gelfond91classical}.
Some well-known recent proposals are presented in \cite{updcaus} and \cite{ap02}
accompanied by very good reviews with many references.

In particular, with the purpose of obtaining a
reliable update process for providing agents with advanced reasoning capabilities, the
authors of \cite{updcaus} present an approach in which new information is incorporated into the current knowledge
base subject to a causal rejection principle.
Causal rejection enforces that in case of
conflicts between rules, more recent rules are preferred and older rules are
overridden \cite{updcaus}. As was noted in \cite{updcaus}, causal rejection is not novel,
and what they provided was a simple and rigorous realization of this principle, as well as the formalization
of a semantics for updating logic programs based on it. We developed the present
work based on this principle also, aiming to express program updates succinctly while preserving
relevant properties.

In our work we analyze the approach presented in \cite{updcaus} because under this approach a program
is explicitly constructed. We believe that this is important from a computational perspective and also that
such an approach is more suitable to study the properties of updates based on the causal rejection principle,
since it provides insight into the operation of the principle itself. We go a step further by presenting an
alternative but equivalent definition of update sequences as presented in \cite{updcaus}, for the case when
considering two programs and that avoids the need
of new atoms. We consider our definition to be more concise and accessible.

In this paper we consider many of the properties presented in \cite{updcaus} because there the authors present
a comprehensive analysis about the properties that an update operator should have.
Moreover, we present new properties for update operators
through an extensive analysis and making use of the new characterization of ASP
by intermediate logics \cite{LiPeVa01,lopstr01,tplp04}.
In this paper, we consider properties similar to the well-known AGM postulates.
We think that it is necessary to reinterpret them in the context of nonmonotonic
reasoning via ASP.
%In addition, we pay particular attention to our view that distinguishes beliefs from justified beliefs.
As a starting point, we study Dalal's Principle \cite{km91} of Independence of Syntax, according
to which the meaning of the knowledge that results from an update
must be independent of the syntax of the original knowledge, as well as independent
of the syntax of the update itself.
In \cite{updcaus} the authors analyze and interpret the AGM postulate corresponding
to Dalal's principle as follows:\\

\hspace*{2.5cm} $ T_1 \equiv T_2 $ implies $ S(K \oplus_f T_1) = S(K \oplus_f T_2)$\\

\noindent where $T_1$ and $T_2$ are any theories, $T_1 \equiv T_2$ denotes that
$T_1$ and $T_2$ are equivalent, i.e. they have the same answer sets.
By $ S(P) $ we denote the set of all
answer sets of $P$, $``\oplus_f"$ is a general update operator.
This interpretation expresses a very demanding principle of independence
of syntax, due to the fact that the AGM postulates were introduced for monotonic logics.
We propose to reconsider the AGM postulates \cite{agm85} under our new interpretation.
%that considers ``justified beliefs" and ``belief".
To this aim we have introduced in
\cite{oz03} a new property, which we call Weak Independence of Syntax, as follows:\\

\hspace*{1.5cm} (WIS): $ T_1 \equiv_{N_2} T_2 $ implies $ S(K \oplus_f T_1) = S(K \oplus_f T_2).$\\

where $T_1$ and $T_2$ are any theories, $T_1 \equiv_{N_2} T_2$ denotes that $T_1$ and $T_2$ are
equivalent in $\logic{N_2}$, in which case we say that they are equivalent knowledge.
We consider this property to be desirable for any update operator based on ASP.
We show that the proposal introduced in \cite{updcaus} for updates
satisfies this principle to a great extent. In fact we show that for programs without
tautologies, this principle holds. We should point out that this property
is accepted as much in belief revision as in updates, as it is shown in
\cite{updcaus}.

The main contributions of our paper can be summarized as follows. First, we present a reinterpretation
of the AGM postulates for ASP based on Nelson logics. Second, we present an analysis of how the update
semantics presented in \cite{updcaus} can be expressed in an alternative form without relying on new atoms.
Finally, we analyze the semantics presented in \cite{updcaus} and our equivalent form in terms of our
adaptation of the AGM postulates.

Our paper is structured as follows: in Section \ref{background} we present
the general syntax of clauses, we also provide the definition of answer sets
for logic programs as well as some background on logic, in particular
on $\logic{N_2}$ logic. In addition, we review the AGM postulates and present the definition of update sequences
given in \cite{updcaus}. Next, in Section \ref{properties} we introduce our $\logic{N_2}$ properties motivated by the
AGM postulates. Section \ref{basicdef} presents an approach for update programs characterized by its simplicity.
In Section \ref{altdef} we introduce an alternative but equivalent definition for update programs as presented
in \cite{updcaus} and analyze its properties. Finally, the conclusions are drawn in Section \ref{conclusions}.

\section{Background}\label{background}

In this section we present a brief background on intuitionistic and intermediate logics and introduce
the notation and key concepts to be used in this paper. In addition we discuss the AGM postulates and present the definition
of update sequences given in \cite{updcaus}.

\subsection{Intuitionistic and Intermediate Logics} \label{subsectIlogs}

Intuitionistic Logic was introduced at the beginning of the twentieth century
and it was developed as a formal system around the 30s.
It is often referred to as Heyting's Constructive Logic and here we will
represent it by the symbol $\logic{I}$. It was proposed as an alternative to
standard propositional logic, which we will represent by $\logic{C}$. The main
feature of intuitionistic logic is that in order to be true a proposition must
be constructively verified, i.e. a proof of it must be constructed. For this
reason, intuitionism is usually bound to concepts like proof and knowledge,
rather than to the traditional classical notion of truth.
The set $\calif{H} := \set{\land, \lor, \lthen, \bot}$ denotes the connectives
of intuitionistic logic. All the connectives in $\calif{H}$ are 2-place
connectives, except for the 0-place $\bot$, known as \emph{falsum}. For any
formula $p$, the expression $\lnot p$ will be understood as an abbreviation of
$p \lthen \bot $ and $\top$ is an abbreviation of $\bot \lthen \bot$. In addition,
in this paper we may use the $\leftarrow$ operator instead of the $\lthen$ operator solely as
a notational convenience.

For a well known axiomatic formalization
of $\logic{I}$ see \cite{Men87}. $\logic{I}$ is strictly contained in
classical logic in the sense that all theorems of $\logic{I}$ are also
theorems of $\logic{C}$, but the converse doesn't hold.
Note that many well
known theorems of $\logic{C}$ are not theorems of $\logic{I}$. The best known
examples usually involve negation, like $p \lor \lnot p$, and
$p \liff \lnot \lnot p$, but there are also formulas without negation that
are theorems of $\logic{C}$ and are not provable in $\logic{I}$, for example
$((p \lthen q) \lthen p) \lthen p$ (\emph{Pierce's law}).

By adding supplementary axioms to intuitionistic logic, we obtain the logics that
are usually known in the literature as \emph{intermediate logics}.
All of them contain all of the intuitionistic theorems, and are contained in the
theorems of classical logic. If this containment is strict, they are called
proper intermediate logics. Here we will just use the term \emph{intermediate
logic} or \emph{$\logic{H}$-logic} to refer to any logic that is as least as
strong as $\logic{I}$, but strictly weaker than $\logic{C}$\footnote{In a
slight abuse of notation we will use the same symbol to refer to a formal
logic system as well as to its theorems.}. Intermediate logics
form a lattice in which the supremum of all intermediate logics is the unique
lower cover of $\logic{C}$. Many names for this logic can be found in the
literature, like \emph{Smetanich logic}, the logic of \emph{Here and There}, and G\"odel 3-valued Logic ($\Gi[3]$).

\subsubsection{G\"odel Multivalued Logics}
These logics are defined by generalizing the idea of truth tables and evaluation
functions of classical logic. G\"odel defined the multivalued logics $\Gi$, with
values in $\set{0, 1, \dots, i-1}$, with the following evaluation function $I$:
\begin{itemize}
\item $I(B \lif A) = i-1$ if $I(A) \leq I(B)$ and $I(B)$ otherwise.
\item $I(A \lor B) = \max(I(A), I(B))$.
\item $I(A \land B) = \min(I(A), I(B))$.
\item $I(\bot) = 0$.
\end{itemize}

An interpretation is a function $I\colon \sig \to \set{0, 1, \dots, i-1}$ that
assigns a truth value to each atom in the language. The interpretation of an
arbitrary formula is obtained by propagating the evaluation of each connective as
defined above. Recall that $\lnot$ and $\top$ were introduced as abbreviations
of other connectives.

For a given interpretation $I$ and a formula $F$ we say that $I$ is a
\emph{model} of $F$ if $I(F) = i-1$. Similarly $I$ is a \emph{model} of a
program $P$ if it is a model of each of the formulas contained in $P$. If $F$ is
modeled by every possible interpretation we say that $F$ is a \emph{tautology}.

Notice that $\Gi[2]$ coincides with classical logic.
The 3-valued logic $\Gi[3]$ is particularly useful for some
of our results.

\subsection{Intermediate Logics and Strong Negation Extensions}\label{nelsonlogic}

In \cite{nelson49falsity} Nelson introduces a logic that extends intuitionistic
logic by adding a new negation connective called \emph{strong negation} and
represented by $\lxnot$. Intuitively, the strong negation connective means that
something is known to be false, not only assumed false due to the absence of a
proof. It allows the constructive falsification of a proposition, in the same
sense as intuitionistic logic allows constructive verification. This logic is
given by the standard connectives of intuitionistic propositional logic, $\calif{H}$, plus
the strong negation connective $\lxnot$.

We will denote by $\logic{N}$ the logic obtained by adding the following axiom
schema to Heyting's axiomatization of intuitionistic logic \cite{Men87}
($\alpha \liff \beta$ stands for
$(\alpha \lthen \beta) \land (\beta \lthen \alpha)$).
%\renewcommand\theenumi{\textbf{N\arabic{enumi}}}
%\begin{enumerate}
%%[label=\textbf{N\arabic{*}}.]
%\begin{enumerate}%[label=\textbf{N}.]
%\item
\\
\\ \hspace*{0.5cm} \textbf{A1}: $\lxnot (\alpha \lthen \beta) \liff \alpha \land \lxnot \beta$
%\item
\\ \hspace*{0.5cm} \textbf{A2}: $\lxnot (\alpha \land \beta) \liff \lxnot \alpha \lor \lxnot \beta$
%\item
\\ \hspace*{0.5cm} \textbf{A3}: $\lxnot (\alpha \lor \beta) \liff \lxnot \alpha \land \lxnot \beta$
%\item
\\ \hspace*{0.5cm} \textbf{A4}: $\alpha \liff \lxnot \lxnot \alpha$
%\item
\\ \hspace*{0.5cm} \textbf{A5}: $\lxnot \lnot \alpha \liff \alpha$
%\item
\\ \hspace*{0.5cm} \textbf{A6}: $\lxnot \alpha \lthen \lnot \alpha$ for atomic $\alpha$\\
%\end{enumerate}
%\renewcommand\theenumi{\arabic{enumi}}

Since $\lxnot \alpha \lthen \lnot \alpha$ is a theorem in $\logic{N}$ for any $\alpha$ (not
only atoms)\cite{Dal86,gurev}, the $\lxnot$ connective is called strong negation. These axioms
are given in \cite{vorobev52a,vorobev52b}, where $\logic{N}$ is presented as
an extension of intuitionistic logic $\logic{I}$, i.e. formulas without the
$\lxnot$ connective are theorems of $\logic{N}$ iff they are theorems of
$\logic{I}$.
Moreover, the same axioms \textbf{A1} to \textbf{A6} can be added to any
intermediate logic $\logic{H}$ to obtain a \emph{least strong negation
extension} of $\logic{H}$, which we will denote by $\logic{N(H)}$. Strong
negation extensions of intermediate logics are sometimes called constructive
logics, but here we will just call them $\logic{N}$-logics following
\cite{kra98}. Sometimes, we will refer to the
\emph{intuitionistic fragment} of a $\logic{N}$-logic $\logic{N}$, i.e. the
intermediate logic whose theorems are exactly those of $\logic{N}$ on the
fragment without $\lxnot$. This fragment will be denoted as $\logic{H(N)}$.
Least strong negation extensions of intermediate logics form a lattice that
was deeply studied by Kracht in \cite{kra98}. There he used
algebraic techniques to prove many results, some of which will be used in
this work. Two particular elements of this lattice will be important for the
present work: the logic $\logic{N}$, extension of $\logic{I}$ and lower bound
of the lattice, and the unique upper bound of it, i.e. the extension of
$\logic{G_3}$ which we will denote by $\logic{N_2}$. The logic $\logic{N_2}$
can be obtained by adding the axiom schema $\alpha \vee (\alpha \rightarrow \beta) \vee \neg \beta$ to $\logic{N}$.

A particular feature of \logic{N} is that the defined connective $\liff$ is
not in general congruential. For instance, it is known that
$\sim \neg a \liff a$ is a theorem, however $\sim \sim \neg a \liff \sim a$ is not a theorem.
For the restricted class of formulas where $\sim$ only applies to atoms,
our $\liff$ connective is again congruential as discussed in \cite{magd}. We will assume this
result freely in the rest of the paper.

It is well--known that provability in $\logic{N}$ can be reduced to provability in
intuitionistic logic following a simple syntactic transformation \cite{ra74}. First, strong negation is moved
towards the atoms using axioms \textbf{A1} to \textbf{A5}. Next, for every atom $a$ of the underlying language,
a new atom $a'$ is introduced and all the formulas $\sim$$a$ are
replaced by $a'$. Finally, formulas of the form $a' \rightarrow \neg a$ are added for each of the new
atoms $a'$.
For example, if we want to know if $\sim$$(a\wedge b)\rightarrow$ $\neg b$ is a theorem in $\logic{N}$, we can
just check if $((a'\rightarrow \neg a) \wedge (b'\rightarrow \neg b)) \rightarrow ((a' \vee b') \rightarrow$
$\neg b)$ is a theorem in intuitionistic logic. By abuse of notation we can
understand $\sim$$a$ as a new atom and we only ask if
$((\sim$$a\rightarrow \neg a) \wedge$$(\sim$$b\rightarrow \neg b))$
$\rightarrow$ $(\sim$$a$ $\vee$ $\sim$$b)$ $\rightarrow$ $\neg$$b$ is a theorem in $\logic{I}$.
Similarly, provability in $\logic{N_2}$ can be reduced to provability in $\logic{G_3}$, the
details of this reduction can be found in \cite{magd}. We will assume this
result freely in the rest of the paper.

\subsection{General Notation}\label{subsectNotation}

We will represent the set of connectives of Nelson's logics by
$\calif{N} := \calif{H} \cup \set{\lxnot}$, where $\lxnot$ is an unary
connective. $\logic{N}$-formulas are the formulas constructed as in standard
propositional logic using the connectives in $\calif{N}$.
For any formulas $\alpha$, $\beta$ the formula $\alpha \liff \beta$ is an
abbreviation of $(\alpha \lthen \beta) \land (\beta \lthen \alpha)$. As was mentioned before,
the $\lnot$ connective can be defined  alternatively as $\lnot \varphi := \varphi \lthen \bot$.
Formulas are constructed over a set of atoms $\mathcal{A}$. An \emph{atom} is just a propositional variable. A \emph{$\lxnot$-literal} is either an atom $a$ or the strong negation of an atom, $\lxnot a$.
A \emph{$\lnot$-literal} is either an atom $a$ or the weak negation of an atom,
$\lnot a$. When we use the term \emph{literal} alone, we mean an
$\lxnot$-literal $l$.
For a literal $l$, the \textit{complementary literal}, $\sim$$l$, is $\sim$$a$ if $l$ = $a$,
and $a$ if $l$ = $\sim$$a$, for some atom $a$.
For a set $S$ of literals, we define $\sim$$S$ = 
\{$\sim$$l \mid l\in S$\}. Similarly, for a set $S$ of formulas we define $\neg$$S$ = \{$\neg$$F \mid F\in S$\}. Also, if
$S$ = \{$P_1,...,P_n$\} is a set of formulas, we define $\bigwedge S$ = $F_1 \wedge ... \wedge F_n$ and 
$\bigvee S$ = $F_1 \vee ... \vee F_n$. We denote by 
$Lit_\mathcal{A}$ the set $\mathcal{A}$ $\cup $ $\sim$$\mathcal{A}$ of all literals over a set of atoms $\mathcal{A}$.
In addition, we define
$\overline{S}$ = $S \cup \lnot\{Lit_\mathcal{A}\setminus S\}$.
A theory is just a finite set of formulas. An \textit{rule} is an expression of the form\\

\hspace*{3.0cm} $A \leftarrow B_1, ..., B_m,$ $\lnot$$B_{m+1} , ...,$ $\lnot$$
B_n$ \hspace*{1.5cm} (1)\\

\noindent where $A$ and each $B_i$ are literals, $\top$, or $\bot$.
If $B_1, ..., B_m,$$\lnot$$B_{m+1} , ..., \lnot$$B_n$ is $\top$
then rule (1) is a fact and may be written simply as $A$.
If $A$ is $\bot$ then rule (1) is a constraint, on the other hand, if $A$ is $\top$ then the rule is a tautology.
An {\it extended logic program} (ELP) \cite{gelfond91classical} is a finite set of rules. Usually, $\mathcal{A}$ will be identified with the set
of all atoms occurring in a program $P$, in which case we call it the \textit{signature} of program $P$. 
Every rule can be identified with a corresponding formula of
the form $\beta_1 \wedge \dots \wedge \beta_m \wedge \lnot \beta_{m+1} \wedge \dots \wedge \lnot \beta_n \rightarrow \alpha$,
where $\alpha$ and each $\beta_i$ are literals. Also, a rule $r$ may be abbreviated as $H(r) \lif B(r)$,
where $H(r)$ and $B(r)$ represent the formulas that comprise the head and body of the rule, respectively.

Due to transformations, our programs can surpass the class of extended logic programs. Hence, it is convenient
to define an \textit{augmented logic program} \cite{LiTaTu99} as a finite set of formulas $F$ that adhere to the following CFG:\\

$R\  ::=\  F \rightarrow  l$\\  
\indent $F\  ::=\  (F\  B\  F)\  |\  \lnot F \ |\  \bot\  |\  \top\ |\ l\  $, where $l$ is a literal \\ 
\indent $B\  ::=\  \vee\  |\  \wedge$\\

\noindent For example, the formula $(a\vee \neg\sim b)\rightarrow \sim p$ 
can be derived according to the previous CFG. Parentheses may be omitted following the
usual conventions regarding the precedence of operators and conjunction may
be expressed with the comma operator instead. Henceforth, when we use the term \textit{program} alone,
we mean an augmented logic program.

With respect to update operators, $\oplus_f$ stands for a general update operator, while $\oplus_i$ for some $i$ such that $1 \leq i \leq n$ represents a specific update operator whose definition is introduced.

\subsection{Answer Sets}
\label{answersets}
Following the characterization of answer sets for logic programs in terms of intermediate logics provided by Pearce in \cite{Pea99},
we adopt the following theorem as a definition of an answer set, which is discussed in detail in \cite{OsNaAr:apal04}.

Let us introduce the notation $T \Vdash_{N_2} M$ to stand for the phrase $T$ is consistent and $T \vdash_{N_2} M$.
We also clarify that by a consistent set of literals $M$, we mean that for all literals $l \in M$ we have either $l \in M$ or $\sim$$l \in M$, but not both. Finally, we define the complement of $M$ as $\widetilde{M} = Lit_{\mathcal{A}} \setminus M$.

\begin{theorem}\label{answersetdef}\cite{OsNaAr:apal04,magd}.
Let $P$ be a program over a set of atoms $\mathcal{A}$, and let $M$ be a consistent set of literals such that $M\subseteq Lit_{\mathcal{A}}$.
M is an answer set of $P$ iff $P \cup \neg \widetilde{M} \cup \neg\neg M \Vdash_{N_2} M$.
\end{theorem}

\noindent Our definition yields different results with respect to that given in \cite{gelfond91classical} only in that according to ours inconsistent programs have no answer sets.\\

\noindent In addition we adopt the notion of strong equivalence, which represents a property that is useful in our work.
\begin{theorem}\label{strongequiv}\cite{LiPeVa01}.
For any programs $P_1$ and $P_2$, $P_1 \equiv_{N_2} P_2$ iff
for every program $P$, $P_1 \cup P$ and $P_2 \cup P$ have the same answer sets.
\end{theorem}

\noindent We also take into account the notion of conservative extension, a property in which a program preserves
its original meaning despite the addition of rules.
\begin{definition}\label{consext}\cite{baral03:cambridge}
Let $P$ and $P'$ be a pair of programs over sets of atoms $\mathcal{A}$ and $\mathcal{A'}$ respectively such that $P \subseteq P'$. We say that $P'$ is a 
\emph{conservative extension} of $P$ if the following condition holds:
$M$ is an answer set for $P$ iff there is an answer set $M'$ for $P'$ such that $M = M' \cap Lit_{\mathcal{A}}$. 
\end{definition}

\noindent Finally, we will introduce two additional results. But first we introduce the following
clarification.\\

\noindent We define that a literal $l$ \textit{occurs} in a formula $F$ recursively as follows\\
\\ \hspace*{1.0cm} $l$ occurs in $l$ \hspace*{1.6cm} where $l$ is a literal \\ \hspace*{1.0cm}
$l$ occurs in $\alpha \circ \beta$ \hspace*{1.0cm} iff $l$ occurs in $\alpha$ or $l$ occurs in $\beta$
\\

\noindent where $\alpha$ and $\beta$ are formulas and $\circ$ denotes one of the operators with which 
we construct our formulas as presented in the CFG
introduced in Section \ref{subsectNotation}. The previous definition allows us to resolve possible ambiguities.
For instance, $a$ does not occur in $\sim$$a$ and similarly $\sim$$a$ does not occur in $a$.

We say that a literal occurs under the scope of the $\lnot$ symbol when it is affected
directly by the $\lnot$ symbol, or is part of a formula affected by the $\lnot$ symbol. For example, in the formula
$a \lif \lnot(x \wedge b)$ the literal $x$ occurs under the scope of $\lnot$. Now consider the formula 
$x \lif \lnot(x \wedge b)$. In this formula, the first occurrence of $x$ is not under the scope of $\lnot$ while the
second does occur under the scope of $\lnot$. We are interested in the case when a literal may \textit{only} occur under the 
scope of the $\lnot$ symbol for some formula $F$, and consequently in a program $P$, where the previous condition holds
for all formulas $F$ identified with the rules of $P$.\\

\noindent The first result can be considered of interest by itself \footnote{Essentially the same result was introduced and proved in \cite{fer05b} as the Completion Lemma.}.

\begin{lemma}\label{formercorollary1}

Let $P$ be a program over a set of atoms $\mathcal{A}$ and x a literal that does not occur in $P$.
Let $F$ be a formula over a set of atoms $\mathcal{F}$ such that $Lit_{\mathcal{F}} \subseteq Lit_{\mathcal{A}}$.
Then  $P \cup \set{x \lif F} \equiv P \cup \set{x \liff F}$.
\end{lemma}

\begin{proof}
See \ref{appendixb}.
\end{proof}

\noindent The second result is the following theorem.

\begin{theorem}\label{equivtheorem}
Let $P$ be a program over a set of atoms $\mathcal{A}$ and x a literal that if occurs in $P$, it occurs only
under the scope of the $\lnot$ symbol.
Let $F$ be a formula over a set of atoms $\mathcal{F}$ such that $Lit_{\mathcal{F}} \subseteq Lit_{\mathcal{A}}$ and x does not
belong to  $Lit_{\mathcal{F}}$.
Then $P \cup \set{x \lif F} \equiv P \cup \set{x \liff F}$.
\end{theorem}

\begin{proof} For a set of literals $M$ we denote by $P^{M(x)}$ the program obtained by substituting $x$ by $\top$ if $x \in M$ or by $\bot$ if 
$x \notin M$. The substitution only occurs under the scope of a $\neg$ symbol in $P$.\\

\noindent \hspace*{2.5cm}$M$ is an answer set of $P \cup \{ x \lif F \}$\\
\hspace*{2.5cm}iff $M$ is an answer set of $P^{M(x)} \cup \{ x \lif F \}$\\
\hspace*{2.5cm}(by Lemma \ref{formercorollary1}, $P^{M(x)} \cup \{ x \lif F \} \equiv P^{M(x)} \cup \{ x \liff F \}$)\\
\hspace*{2.5cm}iff $M$ is an answer set of $P^{M(x)} \cup \{ x \liff F \}$\\
\hspace*{2.5cm}iff $M$ is an answer set of $P \cup \{ x \liff F \}$.
\end{proof}
 
\subsection{AGM Postulates}
\label{agmpostulates}
Although it is not possible to characterize a revision exclusively on logical terms, the general properties of its associated revision function can be defined and algorithms to compute it can be developed, at least for some
cases \cite{agm85}. In order to prove if an algorithm is adequate, we need to establish whether its associated function follows the properties that would normally be expected to hold in a belief revision process. In \cite{agm85} several \textit{rationality postulates} are proposed as a standard for analyzing revision functions. These are referred to as the AGM postulates, in honor of their authors, and are described next.
\\
\\ \hspace*{0.5cm} \textbf{K-1}: \hspace*{0.5cm} for any sentence $\phi$ and any belief set $K$, $K \odot \phi$ is a belief set.
\\ \hspace*{0.5cm} \textbf{K-2}: \hspace*{0.5cm} $\phi \in K \odot \phi$.
\\ \hspace*{0.5cm} \textbf{K-3}: \hspace*{0.5cm} $K \odot \phi \subseteq K + \phi$.
\\ \hspace*{0.5cm} \textbf{K-4}: \hspace*{0.5cm} if $ \neg\phi \notin K$ then $K + \phi \subseteq K \odot \phi$.
\\ \hspace*{0.5cm} \textbf{K-5}: \hspace*{0.5cm} $K \odot \phi = K_\bot$ if and only if $ \vdash\neg\phi$.
\\ \hspace*{0.5cm} \textbf{K-6}: \hspace*{0.5cm} if $\vdash\phi\leftrightarrow\psi$ then $K \odot \phi = K \odot \psi$.\\

In the previous statements $\odot$ represents an arbitrary revision operator as discussed in \cite{agm85}
and not specifically based on answer set semantics, $+$ represents the expansion of a belief set, i.e. adding a sentence regardless of the consistency of the new belief set. A revision operator represents a function taking a belief set and a sentence as arguments and producing a belief set as a result. By belief set we mean a set of sentences closed under logical consequences.
The first postulate requires that the output of the revision function must be a belief set.
The second postulate guarantees that the new sentence $\phi$ will be accepted. The third postulate
states that a revision knows at most, as much as the expansion. In the case in which the new information $\phi$
does not contradict the belief set $K$ which means $\neg\phi \notin K$, the \textbf{K-4} postulate guarantees, in conjunction with \textbf{K-3},
that revision will be equivalent to expansion. Because the goal of revision is to produce a new consistent
belief set, postulate \textbf{K-5} establishes that the new belief set has to be consistent, unless the input sentence $\phi$
is inconsistent itself. The content of the input sentence $\phi$ should determine the revision independently of
its syntax. This means belief revisions should be analyzed on the knowledge level and not on the syntactic level,
therefore logically equivalent sentences should lead to equivalent revisions as postulate \textbf{K-6} says.

Postulates \textbf{K-1} to \textbf{K-6} establish the fundamental properties that any belief revision function should satisfy,
for this reason they are called the \textit{basic} set of postulates. Together, these postulates aim to capture the basic properties that should be respected in an ideal form of belief revision.

In \cite{kat92} the authors argue that the AGM postulates are adequate for the incorporation of new information about a static
world, but not for new information about changes in the world caused by an agent. They formalize the latter behavior
by a set of update postulates. However, all of the basic AGM postulates except \textbf{K-1} have direct correspondences in the update postulates presented in \cite{kat92}, hence they can be considered as a core set of postulates for belief revision as well as for update.

\subsection{Update Programs}
\label{secupd}
In \cite{updcaus} an \textit{update sequence} \textit{\textbf{P}} is defined
as a series $(P_1, P_2, ..., P_n)$ of extended logic programs (ELPs).
Let us consider the definition of update sequence given in \cite{updcaus} but only 
for the case of two programs and let us make a slight change of notation. Our proposal can be 
extended to general case ($P_1, P_2, ..., P_n$) in the iterative form as shown in 
\cite{updcaus}. Under certain conditions, which exclude possibilities for local 
inconsistencies, the iterativity property holds \cite{updcaus}. 
\textit{\textbf{P}} is an update sequence over a set of atoms $\mathcal{A}$
\begin{math} $iff$ \end{math} $\mathcal{A}$ represents the set of atoms 
occurring in the rules of the constituting elements $P_i$ of \textit{\textbf{P}} (1 $\leq$ i $\leq$ 2). 
Given an update sequence \textit{\textbf{P}} = ($P_1$, $P_2$) over $\mathcal{A}$, we assume a set 
$\mathcal{A}^*$ extending $\mathcal{A}$ by new, pairwise distinct atoms $rej(r)$ and 
$A_i$, for each $r$ occurring in \textit{\textbf{P}}, each atom $A$ $\in  \mathcal{A}$,  
and each $i$, 1 $\leq$ $i$ $\leq$ 2. We further assume an injective naming function 
$N$($\cdot , \cdot $), which assigns to each rule $r$ in a program $P_i$ a distinguished
name, $N(r, P_i$), obeying the condition $N(r, P_i$) $\not = $  $N(r', P_j$) 
whenever $i \not =  j$. With a slight abuse of notation we shall identify $r$ with 
$N(r, P_i$) as usual. Finally, for a literal $L$, we write $L_i$ to denote the result
of replacing the atomic formula $A$ of $L$ by $A_i$.

\begin{definition}\label{DZach3}\cite{updcaus}
Given an update sequence \textit{\textbf{P}} = $(P_1, P_2)$ over a set of atoms 
$\mathcal{A}$, we define the update program \textit{\textbf{P$_{\oplus_1}$}} = $P_1$ $\oplus_1$ $P_2$ over
$\mathcal{A}^*$
consisting of the following items: \\

\noindent \hspace*{2.2cm}$(i)$  all constraints in $P_1 $ $\cup$ $P_2$; \\
\hspace*{2.1cm}$(ii)$ for each $r$ $\in$ $P_1$; \\
\hspace*{3.2cm} $\left. \begin{array}{ll} L_1 \leftarrow B(r), \lnot rej(r). \hspace*{1.1cm} \\ 
                rej(r) \leftarrow B(r), \sim L_2. \end{array}\right\}$ \hspace*{0.0cm} if $H(r) = L$; \\
\hspace*{2.0cm}$(iii)$ for each $r$ $\in$ $P_2$; \\
\hspace*{3.5cm}$L_2$ $\leftarrow$ $B(r)$. \hspace*{2.9cm} if $H(r) = L$; \\
\hspace*{2.1cm}$(iv)$ for each literal $L$ occurring in \textit{\textbf{P}}; \\
\hspace*{3.5cm}$L_1$ $\leftarrow$ $L_2$ \\
\hspace*{3.5cm}$L$ $\leftarrow$ $L_1.$ 
\end{definition}

\noindent This definition relies on the fact that rules in program $P_2$ cannot be rejected, which
allows us to construct a simplified program as was noted in \cite{updcaus}. It is important to note that the update sequence \textit{\textbf{P}} is also an extended logic program.
The answer sets of \textit{\textbf{P}} are defined in terms of the answer sets of
\textit{\textbf{P$_{\oplus_1}$}}.

\begin{definition}\label{DZach4}\cite{updcaus}
Let \textit{\textbf{P}} = $(P_1, P_2)$ be an update sequence over a set of atoms $\mathcal{A}$. Then,
$S \subseteq Lit_\mathcal{A}$ is an update answer set of \textit{\textbf{P}} \begin{math} $iff$ \end{math} 
$S = S'$ $\cap$ $\mathcal{A}$ for some answer set $S'$ of \textit{\textbf{P$_{\oplus_1}$}}. The collection of all 
update answer sets of \textit{\textbf{P}} is denoted by $\mathcal{U}($\textit{\textbf{P}}$)$.
\end{definition}

An important remark in relation to update programs is presented in \cite{updcaus}. Update programs do not satisfy
many of the properties defined in the literature. This is partly explained by the nonmonotonicity
of logic programs and the causal rejection principle embodied in the semantics, which strongly depends 
on the syntax of rules \cite{updcaus}. \\

\section{Properties}\label{properties}
In this section we present some basic properties about update operators. The first four are taken from \cite{updcaus}
and are all satisfied by the $\oplus_1$ operator.
The remaining are adapted versions of the AGM postulates. In the statements that follow $\oplus_f$ denotes
an arbitrary update operator and \textit{P} a program. Given two programs $P$ and $P'$, we write $P \equiv P'$ if
they have the same answer sets.\\

\textbf{Initialization:} $\emptyset \oplus_f P \equiv P$ \\

\noindent The authors of \cite{updcaus} present this property as follows: this property
states that the update of an initial empty knowledge base yields just the
update itself.\\

\textbf{Idempotence:} $P \oplus_f P \equiv P$\\

\noindent This property means that the update of program $P$ by itself has no effect \cite{updcaus}.\\

\textbf{Noninterference:} If $P_1$ and $P_2$ are programs defined over disjoint alphabets,
then $P_1 \oplus_f P_2 \equiv P_2 \oplus_f P_1$\\

\noindent This property implies that the order of updates which do not interfere with each other is immaterial \cite{updcaus}.\\

\textbf{Augmented update:} If $P_1\subseteq P_2$ then $P_1 \oplus_f P_2 \equiv P_2$\\

\noindent Updating with additional rules makes the previous update obsolete \cite{updcaus}. Note
that the idempotence property represents a special case of this property.\\

The following properties are motivated by the AGM postulates.
We observe that the interpretation given in \cite{updcaus} of the AGM
postulates is sometimes very demanding, particularly
because the AGM postulates were introduced for monotonic logics.
However, since answer set programming is related to the well known $\logic{N_2}$
monotonic logic, it makes sense to consider properties that relate the monotonic
part of answer sets to its non-monotonic side. That is the objective
of the following properties.\\

Let $P, P_1, P_2,$ and $R$ be programs and $x$ a rule.\\

\noindent
\hspace*{0.7cm} \textbf{BK-1}: \hspace*{0.5cm} $P \oplus_f \{x\} $ is a program.\\
\hspace*{0.7cm} \textbf{BK-2}: \hspace*{0.5cm} $ P_1 \oplus_f P_2 \vdash_{N_2} P_2. $ \\
\hspace*{0.7cm} \textbf{BK-3}: \hspace*{0.5cm} $ P_1 \cup P_2 $ $ \vdash_{N_2} P_1 \oplus_f P_2.$\\
\hspace*{0.7cm} \textbf{BK-4}: \hspace*{0.5cm} if $P_1 \cup P_2$ has answer sets then $P_1 \cup P_2 \equiv P_1 \oplus_f P_2$.\\
\hspace*{0.7cm} \textbf{BK-5}: \hspace*{0.5cm} $ P_2 \equiv_{N_2} \perp $ \begin{math} $ implies
$ \end{math} $ (P_1 \oplus_f P_2) \equiv_{N_2} \perp. $ \\
\hspace*{0.7cm} \textbf{BK-6}: \hspace*{0.5cm} if $ P_1 \equiv_{N_2} P_2 $ then
$ P \oplus_f P_1 \equiv_{} P \oplus_f P_2 $.\\

In the following we call these properties the \textbf{BK-ASP} properties.
In Section \ref{agmpostulates} we restated the 6 principles introduced in \cite{agm85} (\textbf{K-1},..., \textbf{K-6}),
we now translate principle \textbf{K-i} from \cite{agm85} to property \textbf{BK-i} in our approach. However, it should be
noted that this is not the only translation possible.
In \textbf{BK-1} the notion of belief set is replaced by logic program. In the AGM
postulates a sentence is part of a belief set whenever the belief set logically
entails the sentence, for \textbf{BK-2} and \textbf{BK-3} we require that a program has to be inferred
in $\logic{N_2}$. Like their counterparts, together properties \textbf{BK-3} and \textbf{BK-4} state that program update
should be equivalent to expansion whenever possible, only that equivalence is understood
as having the same answer sets. \textbf{BK-5} establishes that an inconsistent program in $\logic{N_2}$
will lead to an inconsistent update. We adopt only one of the implications of the original \textbf{K-5} postulate;
because we consider excessive, in our context, to demand an update operator to derive from an inconsistent program, a new consistent program given a possibly completely independent update\footnote{Consider for instance updating the program $\{a, \leftarrow a\}$ with $\{b\}$.}. \textbf{BK-6} says that equivalent programs should lead to equivalent updates, but the programs have to be equivalent in $\logic{N_2}$, i.e., 
strongly equivalent and it is only expected that the updates will have the same answer sets. Finally, we present
an additional property and prove that it is in fact equivalent to \textbf{BK-6}.

\begin{theorem}\label{bk0}
An update operator $\oplus_f$ satisfies the \textbf{BK-6} (WIS) property if and only if it satisfies the following property\\

\noindent\hspace*{0.7cm} \textbf{BK-0}: \hspace*{0.5cm} if $P_2 \vdash_{N_2} R$ then
$P_1 \oplus_f P_2\equiv_{} P_1 \oplus_f (P_2 \cup R)$.
\end{theorem}

\begin{proof}
To prove the right implication we have that\\
\hspace*{1.5cm}$P_2 \vdash_{N_2} R$ implies $P_2 \equiv_{N_2} P_2 \cup R$\\
now by applying \textbf{BK-6} we obtain\\
\hspace*{1.5cm}$P_1 \oplus_f P_2\equiv_{} P_1 \oplus_f (P_2 \cup R)$ as desired.\\
For the left implication, by applying \textbf{BK-0} we obtain\\
\hspace*{1.5cm}$P$  ${\oplus_f}$ $ P_1 $  ${\equiv}$ $ P $ ${\oplus_f}$ 
                 ( $P_1 \cup P_2$ )\\
similarly,\\
\hspace*{1.5cm}$P$  ${\oplus_f}$ $ P_2 $  ${\equiv}$ $ P $ ${\oplus_f}$ ( $P_2 \cup P_1$ ) \\ 
\noindent
hence\\ 
\hspace*{1.5cm}$P$  ${\oplus_f}$ $ P_1 $  ${\equiv}$ $ P $ ${\oplus_f}$ 
$ P_2 $ as desired.
\end{proof}

\section{A Basic Definition for Updates}\label{basicdef}
In this section we present a simpler new definition for updates. We also prove that it is equivalent to
$\oplus_1$ for a class of programs involving tautologies, which we will define
as \textit{tau-comp}. The relevance of this approach relies on its simplicity and the insight it provides
on the $\oplus_1$ operator, which will become clearer when we introduce a possible refinement of this new operator
as well as our alternative definition of $\oplus_1$.
The notational conventions presented in Section \ref{secupd} apply to this section as well.

\begin{definition}
\label{deftautcomp}
Given an update sequence \textit{\textbf{P}} = $(P_1, P_2)$ over a set of atoms $\mathcal{A}$, we define 
the update program \textit{\textbf{P$_{\oplus_2}$}} = $P_1 \oplus_2 P_2$ over $\mathcal{A}^*$ consisting of the following items: \\

\noindent \hspace*{2.2cm}$(i)$  all constraints in $P_1 \cup P_2$; \\
\hspace*{2.1cm}$(ii)$ for each $r$ $\in$ $P_1$; \\
\hspace*{2.6cm} $L$ $\leftarrow$ $B(r), \lnot\sim$$L$. \hspace*{1.0cm} if $H(r) = L$;\\ 
\hspace*{2.0cm}$(iii)$ all rules $r$ $\in$ $P_2$. \\
\end{definition}

\begin{example}
\label{firstexample}
To illustrate this definition consider the following update sequence.\\

\noindent Let $P_1$ be:
\hspace*{0.81cm} sleep $\leftarrow$ night, $\lnot$watch-tv, $\lnot$other.\\
\hspace*{2.5cm} night.\\
\hspace*{2.5cm} tv-on $\leftarrow$ $\lnot$tv-broke.\\
\hspace*{2.5cm} watch-tv $\leftarrow$ tv-on.\\

\noindent Let $P_2$ be:
\hspace*{0.81cm} $\sim$tv-on $\leftarrow$ power-failure. \\
\hspace*{2.5cm} $\sim$tv-on $\leftarrow$ assignment-due, working. \\
\hspace*{2.5cm} assignment-due.\\
\hspace*{2.5cm} working.\\
\hspace*{2.5cm} other $\leftarrow$ working.\\

If we update $P_1$ with $P_2$ we obtain the following program $P_1$ $\oplus_2$ $P_2$:\\

\hspace*{2.17cm} sleep $\leftarrow$ night, $\lnot$watch-tv, $\lnot$other, $\lnot\sim$sleep.\\
\hspace*{2.5cm} night $\leftarrow$ $\lnot\sim$night.\\
\hspace*{2.5cm} tv-on $\leftarrow$ $\lnot$tv-broke, $\lnot\sim$tv-on.\\
\hspace*{2.5cm} watch-tv $\leftarrow$ tv-on, $\lnot\sim$watch-tv.\\
\hspace*{2.5cm} $\sim$tv-on $\leftarrow$ power-failure. \\
\hspace*{2.5cm} $\sim$tv-on $\leftarrow$ assignment-due, working. \\
\hspace*{2.5cm} assignment-due.\\
\hspace*{2.5cm} working.\\
\hspace*{2.5cm} other $\leftarrow$ working.\\

This program has a unique expected answer set, 
namely\\\\
\hspace*{0.5cm}\{night, other, assignment-due, working, $\sim$tv-on\}.\\

\end{example}

\begin{definition}
We say that a program $P$ is tau-comp w.r.t. a signature $\mathcal{A}$ if every rule of the form  
$l \leftarrow l$ belongs to $P$, where $l$ is a literal over $\mathcal{A}$.
\end{definition}

\noindent We will now prove that $\oplus_1$ and $\oplus_2$ are equivalent for tau-comp programs.
The proof is based on results from \cite{lopstr01,tplp04,Pea99:negation}.

\begin{theorem}
\label{theorem-tautcomp}
Let $P_1$ and $P_2$ be extended logic programs, if $P_2$ is tau-comp then $P_1 \oplus_1 P_2 \equiv P_1 \oplus_2 P_2.$
\end{theorem}

\begin{proof} The idea of the proof is to apply transformations to $P_1 \oplus_1 P_2$ 
with respect to answer sets over the $\mathcal{A}$ signature, to obtain $P_1 \oplus_2 P_2.$
At some point we may delete rules involving a particular literal that we regard as temporary, i. e., a literal
that does not form part of the $\mathcal{A}$ signature of the programs in the update sequence.
This is due to the fact that the program that contains such rules is a conservative extension
of the program that excludes them.\\

Consider a particular positive literal $L$ (i.e. a literal consisting of an atom $a$ not preceded by $\sim$). By construction of $P_1 \oplus_1 P_2$ (defined in Point 
(\textit{iv}) of Definition \ref{DZach3} given in Section \ref{secupd}) the program includes the formulas 
$L_1 \leftarrow L_2$ and $L \leftarrow L_1$ for the literal $L$. \\

\noindent Also, since $P_2$ is tau-comp then it includes the rules $L_2 \leftarrow L$ for the literal $L$. \\

\noindent Hence, $P_1 \oplus_1 P_2 \vdash_{N_2} L_1 \leftrightarrow L_2$ and $P_1 \oplus_1 P_2 \vdash_{N_2} L_2 \leftrightarrow L$. \\

So we can replace each of the $L_1$ and $L_2$ literals by $L$ in the rest of the program as discussed in \cite{ra74,magd}
\footnote{For a positive literal consisting of an atom $a$ we also replace the atoms of the corresponding negative literals $\sim$$a$.}. Then, we can eliminate rules in Point (\textit{iv}) of Definition \ref{DZach3} given previously
as well as rules of the form $L_2 \leftarrow L$ 
($L_1$ and $L_2$ are temporary literals). \\

\noindent So, the rules become \hspace*{0.7cm} $L \leftarrow B(r),$ $\lnot$$rej(r)$ \hspace*{0.6cm}and \\
\hspace*{4.0cm} $rej(r)$ $\leftarrow B(r),$ $\sim$$L$	\hspace*{0.52cm}corresponding to program $P_1.$\\
\hspace*{4.0cm} $L \leftarrow B(r)$	\hspace*{1.97cm}corresponding to program $P_2$. \\

Now, observe that by Theorem \ref{equivtheorem} the answer sets of the program are the same if we replace 
$rej(r) \leftarrow B(r),$ $\sim$$L$ by $rej(r) \leftrightarrow (B(r) \wedge \sim$$L)$. Observe that
this transformation creates a rule whose form is different from that of rules of ELPs, however this 
situation is already considered in \cite{tplp04,Pea99:negation,magd}. Then we can replace $rej(r)$ in $L \leftarrow$ 
$B(r) \wedge$ $\lnot$$rej(r)$ to obtain:	\\
\hspace*{4.0cm} $L \leftarrow B(r) \wedge \lnot$$(B(r) \wedge \sim$$L)$ \hspace*{3.4cm} (*)\\

\noindent then we can delete the rule $rej(r) \leftrightarrow (B(r) \wedge \sim$$L)$ since 
$rej(r)$ is a temporary literal. Finally, to obtain the desired program, the rule of the form (*) can be replaced by 
$L \leftarrow B(r), \lnot\sim$$L$ following the next transformation\\
\hspace*{4.0cm} $L \leftarrow B(r) \wedge (\lnot B(r) \vee \lnot\sim$$L)$ \\\\
by applying the De Morgan law, and\\
\hspace*{4.0cm} $L \leftarrow (B(r) \wedge$ $\lnot$$B(r)) \vee (B(r) \wedge$ $\lnot\sim$$L)$ \\\\
by applying the distributive property. The previous rule can in turn be divided into two rules\\
\hspace*{4.0cm} $L \leftarrow (B(r) \wedge \lnot$$B(r))$ \hspace*{3.43cm} (1)\\
\hspace*{4.0cm} $L \leftarrow (B(r) \wedge \lnot\sim$$L)$ \hspace*{3.54cm} (2)\\\\
Rule (1) is a tautology and can therefore be eliminated, leaving only rule (2) which has the desired form
according to Definition \ref{deftautcomp}.
This process can be applied analogously for the 
rest of the literals. In this way, except for the tautologies of $P_2$, which have no effect on the
answer sets of the derived program, we reach program $P_1 \oplus_2 P_2$ as desired.
\end{proof}

\begin{theorem} \label{teo2} The update operator $\oplus_2$ satisfies the properties \textbf{BK-1}, \textbf{BK-2},
\textbf{BK-3}, \textbf{BK-5}, and \textbf{BK-6}. \\

\begin{proof}
Properties \textbf{BK-1}, \textbf{BK-2}, \textbf{BK-3}, and \textbf{BK-5} follow
directly by construction. The \textbf{BK-6} property follows straightforward by construction and Theorem \ref{strongequiv}.
\end{proof}
\end{theorem}

This allows us to formalize our first result regarding the satisfability of the WIS property for the $\oplus_1$ operator.

\begin{corollary} Let $P$ be an ELP program and let $P_1$ and $P_2$ be tau-comp ELPs. \\
If $P_1 \equiv_{N_2} P_2$  then $P \oplus_1 P_1 \equiv P \oplus_1 P_2.$\\

\begin{proof}
The result is immediate from the equivalence of the $\oplus_1$ and $\oplus_2$ operators
for tau-comp programs presented in Theorem \ref{theorem-tautcomp}.
\end{proof}
\end{corollary}

The \textbf{BK-4} property is not satisfied in general. Consider the following example inspired from \cite{bab03}.

\begin{example}
\label{alf}
\hspace*{1.8cm} Let $P_1$ be: 
\hspace*{0.79cm} day $\leftarrow$ $\lnot$night.\\
\hspace*{4.4cm} night $\leftarrow$ $\lnot$day.\\
\hspace*{4.4cm} see-stars $\leftarrow$ night, $\lnot$cloudy.\\
\hspace*{4.4cm} $\sim$see-stars.
\end{example}

The only answer set of this program is \{$\sim$see-stars, day\}. Now, suppose that $P_1$ is updated with:\\
\hspace*{1.8cm} Let $P_2$ be: 
\hspace*{0.79cm} see-stars $\leftarrow$ see-venus.\\
\hspace*{4.4cm} see-venus $\leftarrow$ see-stars.\\

Applying the $\oplus_2$ operator the answer sets of $P_1 \oplus_2 P_2$ are:
\{see-stars, see-venus, night\}, \{$\sim$see-stars, day\}, and \{$\sim$see-stars, night\}, while $P_1 \cup P_2$ has the
unique answer set \{$\sim$see-stars, day\}. As we can see, the update process
results in additional answer sets, so \textbf{BK-4} doesn't hold.

This proposal satisfies the initialization, idempotence, and augmented update properties.
The initialization property follows directly by construction. The augmented update property
is satisfied because $P_1 \oplus_1 P_2$ is equivalent in \logic{N_2} to $P_2$ whenever $P_1 \subseteq P_2$.
Since the idempotence property is actually a special case of the augmented update property, it holds also.
As the following example shows, this new update operator ($\oplus_2$) doesn't satisfy the 
noninterference property.

\begin{example}
This example shows that the noninterference property is not satisfied by the $\oplus_2$ operator.\\

\hspace*{1.5cm} Let $P_1$ be:
\hspace*{0.74cm} day $\leftarrow$ $\lnot$night.\\
\hspace*{4.4cm} night $\leftarrow$ $\lnot$day.\\
\hspace*{4.4cm} see-stars $\leftarrow$ night, $\lnot$cloudy.\\
\hspace*{4.4cm} $\sim$see-stars.\\

\hspace*{1.5cm} Let $P_2$ be:
\hspace*{0.74cm} $\sim$tv-on $\leftarrow$ power-failure. \\
\hspace*{4.4cm} power-failure.\\

If we update $P_1$ with $P_2$ we obtain the following results:\\

$P_1 \oplus_2 P_2 \not = P_2 \oplus_2 P_1$, because $P_2 \oplus_2 P_1 $ has only one answer set, while
$P_1 \oplus_2 P_2$ adds more models. Thus, the update operator $\oplus_2$ doesn't 
satisfy the noninterference property.
\end{example}

With a slight modification in the definition of $\oplus_2$, this property can be satisfied.
We simply need to add the weakly negated literals only to those rules $r$
in program $P_1$ for which there is a conflictive rule $r'$ in $P_2$, i.e., rules for which $H(r) = \sim H(r')$.
We denote this modified update operator by $\oplus_{2}'$. However, this results in the loss of the \textbf{BK-6}
property, as the following example shows.

\begin{example}\label{exam-wis}
Consider the following programs.\\
\noindent \hspace*{0.5cm} Let $P$ be: \hspace*{0.4cm} $a \leftarrow \lnot\sim$$a$. 
\hspace*{0.5cm} Let $ P_1$ be: \hspace*{0.4cm} $a \leftarrow a$.
\hspace*{0.5cm} Let $ P_2$ be: \hspace*{0.4cm} $b$.\\
\hspace*{2.6cm} $\sim$$a$. 
\hspace*{3.8cm} $b$.
\end{example}
Here $P_1$ and $P_2$ are equivalent in $\logic{N_2}$.
For the update program $P \oplus_{2}' P_1$ we obtain the answer sets $\{a, b\}$ and $\{\sim$$a, b\}$, 
while $P \oplus_{2}' P_2$ has the unique answer set $\{\sim$$a, b\}$, therefore \textbf{BK-6} (WIS) fails.
In addition, the \textbf{BK-4} property is still not satisfied, as applying the $\oplus_{2}'$ operator to the
update sequence of Example \ref{alf} would show. In the next section we present Example \ref{exam-operators}, where $\oplus_{2}'$ also yields unintended answer sets.

\section{An Alternative Definition for Updates}\label{altdef}
In this section we introduce an alternative definition for updates as defined by the $\oplus_1$ operator,
prove that it is equivalent for single updates (i.e., update sequences consisting of two programs), and analyze its properties.
\begin{definition}
\label{operator3}
Given an update sequence \textit{\textbf{P}} = $(P_1,P_2)$ over a set of atoms $\mathcal{A}$, we define 
the update program \textit{\textbf{P$_{\oplus_3}$}} = $P_1 \oplus_3 P_2$ over $\mathcal{A}^*$ consisting of the following items: \\

\noindent \hspace*{2.2cm}$(i)$  all constraints in $P_1 \cup P_2$; \\
\hspace*{2.1cm}$(ii)$ for each $r \in P_1$; \\
\hspace*{2.6cm} $L \leftarrow B(r),$ $\lnot$\textit{\textbf{sup}}$(comp(L), P_2)$. \hspace*{1.0cm} if $H(r) = L$;\\ 
\hspace*{2.0cm}$(iii)$ all rules $r \in P_2$. \\
where\\
\hspace*{2.0cm}$comp(L) =$ $\sim$$L$.\\
\hspace*{2.0cm}$comp(\sim$$L) = L$.\\

\noindent Also, for a program $P$ and a literal $L$ let\\

\noindent \hspace*{2.0cm}$K_L$ = \{ $B(r) \mid H(r)$ = $L, r \in P$ \}\\

\noindent and\\

\begin{equation*} \textit{\textbf{sup}(L,P)} = 
\begin{cases} \bigvee K_L & \text{if $K_L \neq \emptyset$, $\top \notin K_L$.}\\
\bot & \text{if $\lnot\exists$ $r \in P \mid H(r) = L$.}\\
\top & \text{if $\exists$ $r \in P \mid B(r) = \top$, $H(r) = L$.}\\
\end{cases}
\end{equation*}
\\
\end{definition}

Note that \textit{\textbf{sup}(L,P)} essentially takes a disjunction of the bodies of the rules that
have $L$ in their head, or evaluates to $\bot$ in case $L$ does not appear in any of the heads of the rules
of program $P$. Also note that \textit{\textbf{sup}(L,P)} evaluates to $\top$ if $L$ appears as a
fact in $P$, i.e., a rule of the form $L \lif \top$.

\begin{example}
We apply this definition to the update sequence presented in Example \ref{firstexample}.
If we update $P_1$ with $P_2$ we obtain the following program $P_1$ $\oplus_3$ $P_2$:\\

\hspace*{0.68cm} sleep $\leftarrow$ night, $\lnot$watch-tv, $\lnot$other.\\
\hspace*{1.0cm} night.\\
\hspace*{1.0cm} tv-on $\leftarrow$ $\lnot$tv-broke, $\lnot$(power-failure $\vee$ (assignment-due $\wedge$ working))\\
\hspace*{1.0cm} watch-tv $\leftarrow$ tv-on.\\
\hspace*{1.0cm} $\sim$tv-on $\leftarrow$ power-failure. \\
\hspace*{1.0cm} $\sim$tv-on $\leftarrow$ assignment-due, working. \\
\hspace*{1.0cm} assignment-due.\\
\hspace*{1.0cm} working.\\
\hspace*{1.0cm} other $\leftarrow$ working.\\

This program has the same unique answer set that was obtained in Example~\ref{firstexample}, 
namely \{night, other, assignment-due, working, $\sim$tv-on\}.\\

\end{example}

It is important to consider that the updated program is no longer an extended logic program. 
The following lemma shows that our approach is equivalent to the $\oplus_1$ operator.
\begin{theorem}
\label{opequiv}
For any extended logic programs $P_1$ and $P_2$, $P_1 \oplus_1 P_2 \equiv P_1 \oplus_3 P_2.$
\end{theorem}

\begin{proof} We apply a series of transformations to $P_1 \oplus_1 P_2$ 
with respect to answer sets over the $\mathcal{A}$ signature, to obtain $P_1 \oplus_3 P_2$.
At some point we may delete rules involving a particular literal that we regard as temporary, i. e., a literal
that does not form part of the $\mathcal{A}$ signature of the programs in the update sequence.
This is due to the fact that the program that contains such rules is a conservative extension
of the program that excludes them.\\

\noindent \textbf{Note:} in our proof we will apply the following transformation several times. For a subset of rules
$\{L \leftarrow \alpha_1, \dots, L \leftarrow \alpha_n\}$ of a program $P$, where $L$ denotes a particular literal
and each $\alpha_i$ an arbitrary formula representing the body of its respective rule, we join
the rules of the subset into a single rule of the form $L \leftarrow \alpha_1 \vee \dots \vee \alpha_n$ 
by taking the disjunction of the bodies of all the rules.
Thus, such subset of rules can be replaced by a single rule in program $P$, as long as all the rules
that have $L$ as its head have been taken into consideration. This transformation can also be applied
inversely to decompose a rule.\\

\noindent Considering a particular literal $L$, the rules of $P_1 \oplus_1 P_2$ are of the form\\\\
\hspace*{4.0cm} $L_1 \leftarrow B(r)$, $\lnot rej(r)$  \hspace*{1.0cm}*\\
\hspace*{4.0cm} $rej(r) \leftarrow B(r)$, $\sim$$L_2$\\
\hspace*{4.0cm} $L_2 \leftarrow B(r')$\hspace*{2.4cm}*\\
\hspace*{4.0cm} $L_1 \leftarrow L_2$\\
\hspace*{4.0cm} $L \leftarrow L_1$\\\\
where $r$ represents a rule from program $P_1$, $r'$ a rule from program $P_2$, and * denotes the possible existence of multiple rules of the same form (i.e. rules with
the same head but different bodies). The first two types of rules come from Point (\textit{ii}) of Definition \ref{DZach3},
while the third comes from Point (\textit{iii}) and the last two from Point (\textit{iv}). Constraints are omitted since they
undergo no changes.
First, we apply the \textit{unfolding} transformation (see \cite{baral03:cambridge}).
The rule $L_1 \leftarrow L_2$ is unfolded with respect to the rules $L_2 \leftarrow B(r')$, i.e., the rule
$L_1 \leftarrow L_2$ is replaced by $n$ rules of the form $L_1 \leftarrow B_1(r')$ ; ... ; $L_1 \leftarrow B_n(r')$
resulting from each of the $n$ rules $L_2 \leftarrow B_1(r')$ ; ... ; $L_2 \leftarrow B_n(r')$. Thus
the following equivalent set of rules is derived\\\\
\hspace*{4.0cm} $L_1 \leftarrow B(r)$, $\lnot$$rej(r)$  \hspace*{1.0cm}*\\
\hspace*{4.0cm} $rej(r) \leftarrow B(r),$ $\sim$$L_2$\\
\hspace*{4.0cm} $L_2 \leftarrow B(r')$\hspace*{2.4cm}*\\
\hspace*{4.0cm} $L_1 \leftarrow B(r')$\hspace*{2.4cm}*\\
\hspace*{4.0cm} $L \leftarrow L_1$\\\\
By Theorem \ref{equivtheorem}
the answer sets of the program are the same if we replace the rule
$rej(r) \leftarrow B(r)$, $\sim L_2$ by $rej(r) \leftrightarrow (B(r) \wedge \sim$$L_2$).
Then we can replace $rej(r)$ in all rules $L_1 \leftarrow 
B(r) \wedge$ $\lnot$$rej(r)$ to obtain rules of the form	\\
\hspace*{4.0cm} $L_1 \leftarrow B(r) \wedge \lnot(B(r) \wedge\sim$$L_2$) \hspace*{2.2cm} (1)\\\\
\noindent and afterwards we can delete the rule $rej(r) \leftrightarrow (B(r) \wedge \sim$$L_2$) since 
$rej(r)$ is a temporary literal. The rule of the form (1) can be replaced by 
$L_1 \leftarrow B(r) \wedge \lnot\sim$$L_2$ following the next transformation. First, by applying the De Morgan law we obtain\\
\hspace*{4.0cm} $L_1 \leftarrow B(r) \wedge (\lnot B(r) \vee \lnot\sim$$L_2)$ \\\\
Then, by the distributive property the previous rule becomes\\
\hspace*{4.0cm} $L_1 \leftarrow (B(r) \wedge \lnot B(r)) \vee (B(r) \wedge \lnot\sim$$L_2)$ \\\\
This rule can in turn be divided into two rules\\
\hspace*{4.0cm} $L_1 \leftarrow (B(r) \wedge \lnot B(r))$ \hspace*{3.3cm} (a)\\
\hspace*{4.0cm} $L_1 \leftarrow (B(r) \wedge \lnot\sim$$L_2)$ \hspace*{3.25cm} (b)\\\\
Rule (a) is a tautology and can therefore be eliminated. In this way we arrive to the following
set of rules\\\\
\hspace*{4.0cm} $L_1 \leftarrow B(r)$, $\lnot\sim$$L_2$  \hspace*{1.2cm}*\\
\hspace*{4.0cm} $L_2 \leftarrow B(r')$\hspace*{2.43cm}*\\
\hspace*{4.0cm} $L_1 \leftarrow B(r')$\hspace*{2.43cm}*\\
\hspace*{4.0cm} $L \leftarrow L_1$\\\\
Now we join all the rules that contain 
$\sim$$L_2$ in the head into a single rule\\
\hspace*{3.3cm} $\sim$$L_2$ $\leftarrow B_1(r'') \vee B_2(r'') \vee ... \vee B_n(r'')$ \hspace*{2.0cm} (c)\\\\
where $r''$ represents a rule of program $P_2$. By Theorem \ref{equivtheorem} this allows us to replace $\sim$$L_2$ in rules (b) to obtain rules of the form\\
\hspace*{3.0cm} $L_1 \leftarrow B(r), \lnot (B_1(r'') \vee B_2(r'') \vee ... \vee B_n(r''))$ \hspace*{1.20cm} (d)\\\\
Note that rule (d) has the form of Point (\textit{ii}) of our definition and that rule (c) can be
eliminated since $\sim$$L_2$ is a temporary literal.
Now we join the bodies of all rules (d) and $L_1 \leftarrow B(r')$ into a single rule $L_1 \leftarrow \alpha$ where $\alpha$ abbreviates
the composed body. Similarly, we also join the bodies of all the rules $L_2 \leftarrow B(r')$ to obtain the rule
$L_2 \leftarrow \beta$. Thus we derive a new set of rules\\
\hspace*{5.0cm} $L_1 \leftarrow \alpha$\\
\hspace*{5.0cm} $L_2 \leftarrow \beta$\\
\hspace*{5.0cm} $L \leftarrow L_1$\\\\
Then we can delete the rule $L_2 \leftarrow \beta$ since $L_2$ is a temporary literal. Next we replace
$L_1$ in the third rule for $\alpha$ generating the rule $L \leftarrow \alpha$. At this point the rule
$L_1 \leftarrow \alpha$ can be eliminated because $L_1$ is a temporary literal. By decomposing the remaining
rule $L \leftarrow \alpha$ we can generate the rules\\\\
\hspace*{2.0cm} $L \leftarrow B(r), \lnot (B_1(r'') \vee B_2(r'') \vee ... \vee B_n(r''))$ \hspace*{0.5cm}*\\
\hspace*{2.0cm} $L \leftarrow B(r')$\hspace*{5.83cm}*\\\\
The first rule represents the set of rules of the Point (\textit{ii}) of Definition \ref{operator3}, while the second rule represents the set of rules created with Point (\textit{iii}). We can apply this process analogously for the 
rest of the literals. After this sequence of transformations we obtain the same set of rules as by applying our definition, thus the two operators are equivalent.
\end{proof}

Initially, it may appear that the $\oplus_3$ operator could be equivalent to the $\oplus_{2}'$ operator discussed in Section
\ref{basicdef}. However, as the following example shows, this is not the case.

\begin{example}\label{exam-operators}
Consider the following programs.\\
\noindent \hspace*{1.0cm} Let $P_1$ be: \hspace*{0.4cm} $a \leftarrow \lnot\sim$$a$. 
\hspace*{1.0cm} Let $ P_2$ be: \hspace*{0.4cm} $a \leftarrow c$.\\
\hspace*{3.23cm} $\sim$$a$. 
\hspace*{4.3cm} $b$.\\
\hspace*{3.23cm} $\sim$$c$.
\end{example}

The updated program $P_1 \oplus_{2}' P_2$ has the answer sets \{$\sim$a, b, $\sim$c\} and \{a, b, $\sim$c\} while
$P_1 \oplus_3 P_2$ has the unique answer set \{$\sim$a, b, $\sim$c\}. For this particular example, we can say
that the $\oplus_3$ operator incorporates the new knowledge more accurately, since $a$ should not be derived
if it depends on $c$, which is not true in the first place. Nevertheless, for the cases in which there are no
conflicting literals in the first of the programs, the operators in question would be equivalent.\\

Next we will discuss how this definition of update programs relates to the \textbf{BK-ASP} properties presented
earlier.

\begin{definition}
\label{ptau-free}
A program $P$ is tau-free w.r.t. a signature $\mathcal{A}$ if no rule of the form $l \leftarrow l, \alpha$  belongs to $P$,
where $\alpha$ is a formula which could be empty and $l$ a literal over $\mathcal{A}$. By abuse of notation we define
a tau-free rule as a rule that does not follow the structure just described.
\end{definition}

\begin{theorem}\label{TZach0}
The update operator $\oplus_3$ satisfies the \textbf{BK-ASP} properties \textbf{BK-1}, \textbf{BK-2}, \textbf{BK-3}
and \textbf{BK-5} for any extended logic programs $P$, $P_1$, and $P_2$ as well as \textbf{BK-6} if $P_1$ and $P_2$ are tau-free ELPs.
\end{theorem}

\begin{proof} The first four properties follow directly by construction. The proof that \textbf{BK-6} holds for tau-free
ELPs for the operator $\oplus_1$ is presented in Lemma \ref{LZach2} of \ref{appendixa}. Because the operator $\oplus_3$ is equivalent to the operator $\oplus_1$ that result applies as well.
\end{proof}

We now make use of the examples of Section \ref{basicdef} to demonstrate that the properties \textbf{BK-4} and \textbf{BK-6}
are not satisfied in general by the update operator $\oplus_3$ and equivalently by $\oplus_1$.\\

First, we apply the operator $\oplus_1$ to the update sequence in Example \ref{alf}. The answer sets of $P_1 \oplus_1 P_2$ are:
\{see-stars, see-venus, night\} and \{$\sim$see-stars, day\}, while $P_1 \cup P_2$ has the
unique answer set \{$\sim$see-stars, day\}. As we can see, the update process
adds a second answer set, so \textbf{BK-4} doesn't hold.

Now, when we consider the update sequences of Example \ref{exam-wis} we obtain the following results.
The programs $P_1$ and $P_2$ are equivalent in $\logic{N_2}$.
For the update program $P \oplus_1 P_1$ we obtain the answer sets $\{a, b\}$ and $\{\sim$$a, b\}$, 
while $P \oplus_1 P_2$ has the unique answer set $\{\sim$$a, b\}$, therefore \textbf{BK-6} (WIS) fails.\\

In order to resolve the previous discrepancies, the update semantics can be refined in the following way

\begin{equation*}
P_1 \oplus_{3}' P_2 = 
\begin{cases} P_1 \cup P_2 & \text{if $P_1 \cup P_2$ has answer sets}\\
P_1 \oplus_3 P_2' & \text{otherwise}
\end{cases}
\end{equation*}

\noindent where $P_2'$ is $P_2$ without tautologies. In this way all of the \textbf{BK-ASP} properties we have presented are satisfied.

\begin{theorem}\label{TZach02}
The update operator $\oplus_{3}'$ satisfies the \textbf{BK-ASP} properties \textbf{BK-1} to \textbf{BK-6} for any 
extended logic programs.
\end{theorem}

\begin{proof} Because of the refinement just introduced, property \textbf{BK-4} holds and the rest of the first five properties follow directly by construction. We prove that \textbf{BK-6} holds for the operator $\oplus_{3}'$ as follows.
For the ELPs $P_1$ and $P_2$ we assume $P_1 \equiv_{N_2} P_2$. Now if $P \cup P_1$ has answer sets, by Theorem \ref{strongequiv} these are the same answer sets of $P \cup P_2$, which is $P \oplus_{3}' P_2$ according to our definition. If $P \cup P_1$ does not have answer sets, then Lemma \ref{LZach2} (see \ref{appendixa}) applies, since tautologies are removed from the right operand, i.e., the program that introduces the new rules.
\end{proof}

The previous theorem demonstrates that in principle it is possible to define a semantics that satisfies all of the
\textbf{BK-ASP} properties, which is important from a mathematical viewpoint. However, this semantics is not free
of counterintuitive and undesirable behaviors, as was brought to our attention by one of our anonymous reviewers 
with the following examples (adapted to include strong negation).

\begin{example}\label{referee1}
Consider the following programs.\\
\noindent \hspace*{0.5cm} Let $P$ be: \hspace*{0.5cm} see-stars.
\hspace*{1.5cm} Let $P_1$ be: \hspace*{0.5cm} see-stars $\leftarrow$ see-stars.\\
\hspace*{2.7cm} $\sim$see-stars.\\
\\
\hspace*{0.5cm} Let $P_2$ be: \hspace*{0.38cm} see-stars $\leftarrow$ see-venus.\\
\hspace*{2.7cm} see-venus $\leftarrow$ see-stars.
\end{example}

The program $P \oplus_{3}' P_1$ does not have answer sets, as would normally be expected. However, the program
$P \oplus_{3}' P_2$ has the single answer set \{see-stars, see-venus\}, which intuitively should not have
been derived from the new information in $P_2$.

\begin{example}\label{referee2}
Consider the following programs.\\
\noindent \hspace*{0.3cm} Let $P_1$ be: \hspace*{0.4cm} open-school.
\hspace*{1.3cm} Let $P_2$ be: \hspace*{0.4cm} $\sim$open-school $\leftarrow$ holiday.\\
\hspace*{2.5cm} holiday $\leftarrow$ $\lnot$workday.
\hspace*{2.2cm} workday $\leftarrow$ $\lnot$holiday.\\
\\
\hspace*{0.3cm} Let $P_3$ be: \hspace*{0.4cm} see-stars.
\hspace*{1.8cm} Let $P_4$ be: \hspace*{0.4cm} $\sim$see-stars.
\end{example}

Program $P_1 \oplus_{3}' P_2$ has the single answer set \{open-school, workday\}, which is the same
single answer set of $P_1 \cup P_2$. However, program $(P_1 \cup P_3) \oplus_{3}' (P_2 \cup P_4)$ has two
answer sets, namely \{open-school, workday, $\sim$see-stars\} and \{holiday, $\sim$open-school, $\sim$see-stars\}. The addition of rules affects the interpretation
of other rules, despite having disjoint alphabets. Therefore the $\oplus_{3}'$ operator violates the general principle
that completely independent parts of a program should not interfere with each other.

The existence of such undesirable behaviors suggests the need of analyzing alternative semantics for logic program updates,
considering for instance the semantics presented in \cite{dylps} and \cite{homola01}. We consider this to be an interesting direction for future work.

\section{Conclusions}
\label{conclusions}
We have presented a set of properties based on $\logic{N_2}$ logic that provide a reinterpretation the AGM postulates
into the context of answer set programming. These properties hold for update programs following the definition
presented in \cite{updcaus} after an unsubstantial modification of its semantics.
In addition, we introduced an alternative definition
for update programs that is equivalent, yet more concise and intuitive.
There are many aspects to consider in the theory of belief revision and update.
The desired behavior in a particular situation may prove inappropriate in
a different setting. It is essential to characterize the properties that ought to
be respected in the process of belief change, however, it is unlikely that such
characterization can be reduced to a set of generally accepted and applicable
postulates. We can also conclude that the approach based on
$\logic{N_2}$ logic for the characterization of answer sets has proven useful to obtain a better understanding
of the answer set semantics, as well as to study many of its interesting properties.

We emphasize that with the properties presented in this paper we aim only to provide a starting point for the analysis of belief revision under answer set semantics. Still more work needs to be done to identify new interesting properties and define alternative operators for belief change.\\

\noindent \textbf{Acknowledgements}\\

\noindent We would like to thank Vladimir Lifschitz for his remark regarding Lemma \ref{formercorollary1}.

\appendix

\section{Analysis of the WIS property}
\label{appendixa}

We prove that the operator $\oplus_1$ satisfies the 
Weak Independence of Syntax property for tau-free programs.

\begin{definition}\cite{updcaus}\label{DZach5}
Let us call two rules $ r_1 $ and $ r_2 $ conflicting \begin{math} $iff$ \end{math}
$H(r_1) =$ $\sim$$H(r_2)$,
we denote this by $r_1 \bowtie r_2$. 
\end{definition}

\begin{definition}\cite{updcaus}
For an update sequence \textit{\textbf{P}} = $(P,...,P_n)$ over a set of atoms $\mathcal{A}$ and 
$S \subseteq Lit_\mathcal{A}$ based on the principle of founded rule rejection, we define 
the rejection set of $S$ by $Rej(S,$ \textit{\textbf{P}}$)$ $=$ $\cup_{i=1}^n$ $Rej_{i}(S,$ \textit{\textbf{P}}$)$, where 
$Rej_n(S,$ \textit{\textbf{P}}$)$ $=$ $\emptyset$ and, for $n > i \geq 1$,\\

\hspace*{1.1cm}$Rej_i(S,$ \textit{\textbf{P}}$)$ $=$ $\{r\in P_i \mid \exists r'\in P_j \setminus Rej_j(S,$ \textit{\textbf{P}}$)$,
for some $j \in \{i+1, ..., n\}$\\
\hspace*{4.1cm} such that $r \bowtie r'$ and $\overline{S}\vdash_{N_2} B(r) \wedge B(r')\}.$
\end{definition}

\begin{definition}\cite{updcaus}
For an update sequence \textit{\textbf{P}} = $(P,...,P_n)$ over a set of atoms $\mathcal{A}$ and 
$S \subseteq Lit_\mathcal{A}$, let us define\\

\hspace*{1.1cm}$Rej'(S,$ \textit{\textbf{P}}$)$ $=$ $\cup_{i=1}^n\{r\in P_i \mid \exists r'\in P_j$,
for some $j \in \{i+1, ..., n\}$\\
\hspace*{4.1cm} such that $r \bowtie r'$ and $\overline{S}\vdash_{N_2} B(r) \wedge B(r')\}.$
\end{definition}

\noindent\textbf{Note a:} Observe that $Rej$ and $Rej'$ coincide for two programs.
Furthermore $Rej'(S, (P_1, P_2)) = \{r\in P_1 \mid \exists r'\in P_2$, such that 
$r \bowtie r'$, and $\overline{S}\vdash_{N_2} B(r) \wedge B(r')\}$. So $Rej'$ only contains clauses from $P_1$ 
and not from $P_2$, i.e., only rules from $P_1$ are rejected.\\

\noindent We denote by $\cup \textit{\textbf{P}}$ the set of all rules occurring in $\textit{\textbf{P}}$, i.e., 
$\cup \textit{\textbf{P}}$ = $\cup_{i=1}^n P_{i}$. By $P^S$ we denote the standard reduct of a program $P$
with respect to a set of literals $S$.

\begin{lemma}\label{LZach6}
Let \textit{\textbf{P}} = $(P_1, P_2)$ be an update sequence over a set of atoms $\mathcal{A}$ and $S \subseteq Lit_\mathcal{A}$ a set 
of literals. Then, $S$ is an answer set of \textit{\textbf{P}} \begin{math} $iff$ \end{math} $S$ is an answer set of 
$(P_1 \backslash Rej'(S, \textit{\textbf{P}})) \cup P_2$. 
\end{lemma}

\begin{proof} 
%Directly by theorem 4 given in \cite{updcaus} and the previous note. \\
$S$ is an answer set of \textit{\textbf{P}} \begin{math} $iff$ \end{math}\\
$S$ is the minimal model of $(\cup \textit{\textbf{P}}\setminus Rej(S, \textit{\textbf{P}}))^S$ (by Theorem 4 given in \cite{updcaus}) \begin{math} $iff$ \end{math} \\
$S$ is an answer set of $(P_1 \cup P_2) \setminus Rej(S, \textit{\textbf{P}})$ \begin{math} $iff$ \end{math} \\
$S$ is an answer set of $(P_1 \setminus Rej'(S, \textit{\textbf{P}})) \cup P_2$ (by Note a) as desired.
\end{proof}

\begin{lemma}\label{new_lemma}
Let $P$ be an extended logic program, $x$ a literal, and $M$ a set of literals.
If $P\vdash_{N_2} x$ and $\overline{M}\vdash_{N_2} P$,
then $\exists$ $r \in P$ such that $r$ has the form $x\leftarrow \beta$ and $\overline{M}\vdash_{N_2} \beta$.
\end{lemma}

\begin{proof} 
Due to the translation from Nelson logics to intermediate logics, we may suppose to be in $\logic{G_3}$.
The translation adds more rules to the programs but these are constraints, so the result is 
not affected. Additionally, this allows us to replace the literal $x$ by an atom $a$.
Assume by contradiction that $\forall r\in P$ such
that $r$ has the form $a\leftarrow \beta$, $\overline{M}\vdash_{N_2} \beta$ is false.
We need to prove that $P\vdash_{N_2} a$ is false. Let us construct an interpretation $I$ based on 
$\overline{M}$. If an atom belongs to $\overline{M}$ we assign to it the value 2. If
the negation of an atom belongs to $\overline{M}$ we assign to it the value 0. Since 
$\overline{M} \vdash_{N_2} P$ then clearly $I$ models $P$. Now we construct another 
interpretation $I'$ based on $I$ as follows. All atoms but $a$ are assigned the same value
as in $I$, while $a$ is assigned the value 1. It is easy but tedious to check that $I'$
also models $P$. However, $I'$ does not model $a$, as desired.
\end{proof}

\begin{lemma}\label{Lg3}
Let $P_1$, $P_2$, $\{c\}$, and $\{r\}$ be extended logic programs where $r\in P_1$ and $c$ is a tau-free rule.
Let $M$ be a set of literals. 
Suppose $P_2 \vdash_{N_2} c$, $\overline{M} \vdash_{N_2} B(r) \wedge B(c)$, and $\overline{M}\vdash_{N_2} P_2$. Also,
$r$ and $c$ are conflicting rules. Then $\exists$ $r'$$\in$$P_2$ such that
$r'$ is a conflicting rule with $r$ and $\overline{M} \vdash_{N_2} B(r').$
\end{lemma}

\begin{proof} 
Let $c$ and $r$ be formulas of the form $x\leftarrow \theta$ and $\sim$$x\leftarrow \beta$, respectively.
We assume that $r\in P_1$, $\overline{M}\vdash_{N_2} \beta$, and that $c$ and $r$ are conflicting rules. 
Furthermore, we assume $\overline{M} \vdash_{N_2} P_2\cup \{\theta\}$. 
Also by hypothesis $P_2 \vdash_{N_2} \theta \rightarrow x$, therefore $P_2\cup \{\theta\} \vdash_{N_2} x$. 
Then by Lemma \ref{new_lemma}, $\exists$ $ r'\in P_2\cup \{\theta\}$
such that $r'$ is of the form $x  \leftarrow B(r')$ and $\overline{M} \vdash_{N_2} B(r')$.
Moreover $r' \in P_2$, due to the restriction that $c$ is a tau-free rule and consequently
cannot contain $x$ in its body.
\end{proof}

\begin{lemma}\label{LZachnew}
Let $P_1$, $P_2$, and $\{c\}$ be extended logic programs where $\{c\}$ is tau-free.
Suppose $\overline{S}\vdash_{N_2} P_2$ and $P_2 \vdash_{N_2} c$.
Then $Rej'(S,(P_1,P_2))$ = $Rej'(S,(P_1,P_2\cup \{c\})$. \\

\begin{proof} 
We need to prove two cases:\\

Case 1) $Rej'(S,(P_1,P_2)) \subseteq Rej'(S,(P_1,P_2\cup \{c\}).$ Clearly holds. \\

Case 2) $Rej'(S,(P_1,P_2\cup \{c\}) \subseteq Rej'(S,(P_1,P_2))$ \\

\noindent Let $r \in Rej'(S,(P_1,P_2\cup \{c\}))$ then $\exists r'\in (P_2\cup \{c\})$ such that
$\overline{S}\vdash_{N_2} B(r) \wedge B(r').$ Here we have two cases: \\
\hspace*{0.4cm} a) If $r'\in P_2$ then $r\in Rej'(S,(P_1,P_2))$ as desired. \\
\hspace*{0.4cm} b) If $r' = c$, we have $r\in P_1$, $P_2\vdash_{N_2} c$,
$\overline{S} \vdash_{N_2} B(r) \wedge B(c)$, $r$ and $c$ are conflicting rules, and by hypothesis $\overline{S}\vdash_{N_2} P_2$. Then by Lemma \ref{Lg3}, 
$\exists r"\in P_2$ such that $r$ and $r"$ are conflicting rules and 
$\overline{S}\vdash_{N_2} B(r) \wedge B(r")$. Hence,  $r\in Rej'(S,(P_1,P_2))$ as desired.
\end{proof}
\end{lemma}

\noindent \textbf{Note b:} $P' \vdash_{N_2} c$ \hspace*{0.2cm}implies 
\hspace*{0.2cm} $ P' \equiv_{N_2} P' \cup \{c\}. $ 

\begin{lemma}
\label{LZach1}
Let $P_1$, $P_2$, and $\{c\}$ be extended logic programs where $\{c\}$ is tau-free.
If $P_2\vdash_{N_2} c$ then
$P_1 \oplus_1 P_2 \equiv P_1 \oplus_1 (P_2 \cup \{c\}).$ 
\end{lemma}

\begin{proof}
$S$ is an answer set of $P_1 \oplus_1 P_2$ iff \hspace*{5.28cm} by Lemma \ref{LZach6} \\
$S$ is an answer set of $(P_1 \setminus Rej'(S,(P_1,P_2)))\cup P_2$ iff \hspace*{2.33cm} by Lemma \ref{LZachnew}\\
$S$ is an answer set of $(P_1 \setminus Rej'(S,(P_1,P_2\cup \{c\})))\cup P_2$ iff \hspace*{1.46cm} by Note b\\
$S$ is an answer set of $(P_1 \setminus Rej'(S,(P_1,P_2\cup \{c\})))\cup (P_2\cup \{c\})$ iff \hspace*{0.3cm} by Lemma \ref{LZach6}\\
$S$ is an answer set of $P_1 \oplus_1 (P_2\cup \{c\})$ \\

\hspace*{0.0cm} Hence, $P_1 \oplus_1 P_2 \equiv P_1 \oplus_1 (P_2 \cup \{c\})$ as desired.
\end{proof}

Now we can present our main result.

\begin{lemma}\label{LZach2}
Let $P_1$, $P_2$, and $R$ be extended logic programs where $R$ is tau-free.
Suppose that $P_2 \vdash_{N_2} R$,
then
 $P_1$ ${\oplus_1}$ $P_2$ ${\equiv}$ $P_1$ ${\oplus_1}$ $( P_2 \cup R ).$ 
\end{lemma}

\begin{proof}
By induction on the size of $R$. \\ \\
\noindent \textbf{Base case:} Let $R = \emptyset$, the result is immediate.\\
\noindent \textbf{Induction step:} Let $P_2$, $R$, and $\{c\}$ be extended logic programs where $R$ and $\{c\}$ are tau-free. 
We need to show:\\

\hspace*{0.8cm} if $P_2 \vdash_{N_2} R \cup \{c\}$ \hspace*{0.3cm} then \hspace*{0.3cm} 
$P_1 \oplus_1 P_2 \equiv P_1 \oplus_1 (P_2 \cup (R \cup \{c\})).$  \\

\noindent But we know that $P_2 \vdash_{N_2} R \cup \{c\}$ means that 
$P_2 \vdash_{N_2} R$ and $P_2 \vdash_{N_2} c$, then by applying the induction hypothesis\\ 
\hspace*{5.6cm} $P_1 \oplus_1 P_2 \equiv P_1 \oplus_1 (P_2 \cup R) \hspace*{2.4cm}$  (I)\\ 
\noindent By Lemma \ref{LZach1}, if $P_2 \cup R \vdash_{N_2} c$ then 
\hspace*{0.5cm}$P_1 \oplus_1 (P_2 \cup R) \equiv P_1 \oplus_1$ 
                  $((P_2 \cup R) \cup \{c\})$ \hspace*{0.4cm}  (II) \\

\noindent Now, from (I) and (II) we have 
\hspace*{0.26cm} $P_1 \oplus_1 P_2 \equiv P_1 \oplus_1 ((P_2 \cup R) \cup \{c\})$  \\
Since \hspace*{4.37cm}$P_1 \oplus_1 ((P_2 \cup R ) \cup \{ c \})$ 
$\equiv P_1 \oplus_1 (P_2 \cup ( R \cup \{ c \}))$ \\
we obtain \hspace*{3.6cm} $P_1 \oplus_1 P_2 \equiv P_1 \oplus_1 (P_2 \cup (R \cup \{c\}))$ as desired.
\end{proof}

We have proven that the \textbf{BK-0} property holds for the $\oplus_1$ operator if $R$ is tau-free.
If we apply the proof of Theorem \ref{bk0} of Section \ref{properties} under this restriction, this results in the fact
that \textbf{BK-6} holds provided $P_1$ and $P_2$ are tau-free.

\section{Proof of Lemma \ref{formercorollary1}}
\label{appendixb}

Recall that an alternative proof is presented in \cite{fer05b}.\\

\noindent Let $P$ be a program over a set of atoms $\mathcal{A}$ and $x$ a literal that does not occur in $P$.
Let $F$ be a formula over a set of atoms $\mathcal{F}$ such that $Lit_{\mathcal{F}} \subseteq Lit_{\mathcal{A}}$.
Then  $P \cup \set{F \lif x} \equiv P \cup \set{x \liff F}$.\\

\noindent If we apply the definition of an answer set given in Theorem \ref{answersetdef}, then we only need to prove\\

\noindent \hspace*{2.0cm}$P, F \lthen x, \neg \widetilde{M} \cup \neg\neg M \Vdash_{N_2} M$ \hspace*{1.0cm} iff\\
\hspace*{2.0cm}$P, F \liff x, \neg \widetilde{M} \cup \neg\neg M \Vdash_{N_2} M$\\

\noindent First, we apply the transformation described in Section \ref{nelsonlogic} to reduce the previous statement to provability in \logic{G_3}, obtaining\\

\noindent \hspace*{2.0cm}$P', F' \lthen x, \neg(x' \wedge x), C', \neg \widetilde{M} \cup \neg\neg M \Vdash_{G_3} M$ \hspace*{1.0cm} iff\\
\hspace*{2.0cm}$P', F' \liff x, \neg(x' \wedge x), C', \neg \widetilde{M} \cup \neg\neg M \Vdash_{G_3} M$\\

\noindent where $P'$ represents program $P$ after the change of literals $\sim$$l$ by atoms $l'$, $F'$ the formula $F$ after
the same changes, and $C'$ the set of added constraints (excluding the constraint involving $x$). Now we have two cases.\\

\noindent I) $x \notin M$ so $\neg x \in \neg \widetilde{M}$\\

\noindent In this case the statement to prove becomes\\

\noindent \hspace*{2.0cm}$P', F' \lthen x, \neg x, \neg(x' \wedge x), C', \neg \widetilde{M'} \cup \neg\neg M \Vdash_{G_3} M$ \hspace*{1.0cm} iff\\
\hspace*{2.0cm}$P', F' \liff x, \neg x, \neg(x' \wedge x), C', \neg \widetilde{M'} \cup \neg\neg M \Vdash_{G_3} M$\\

\noindent where $\widetilde{M'}$ denotes the set $\widetilde{M'}$ excluding the atom $x$.
It is easy to show that $F' \lthen x, \neg x \equiv_{G_3} F' \liff x, \neg x$ so the statement holds
for both implications.\\

\noindent II) $x \in M$ so $\neg\neg x \in \neg\neg M$\\

\noindent In this case the statement to prove becomes\\

\noindent \hspace*{2.0cm}$P', F' \lthen x, \neg\neg x, \neg(x' \wedge x), C', \neg \widetilde{M} \cup \neg\neg M' \Vdash_{G_3} M$ \hspace*{1.0cm} iff\\
\hspace*{2.0cm}$P', F' \liff x, \neg\neg x, \neg(x' \wedge x), C', \neg \widetilde{M} \cup \neg\neg M' \Vdash_{G_3} M$\\

\noindent where $M'$ denotes the set $M'$ excluding the atom $x$. It is easy to show that this statement can be
reduced to\\

\noindent \hspace*{2.0cm}$P', F' \lthen x, \neg\neg x, \neg x', C', \neg \widetilde{M} \cup \neg\neg M' \Vdash_{G_3} M$ \hspace*{1.0cm} iff\\
\hspace*{2.0cm}$P', F' \liff x, \neg\neg x, \neg x', C', \neg \widetilde{M} \cup \neg\neg M' \Vdash_{G_3} M$\\

\noindent We will abbreviate the statement above in the following way\\

\noindent \hspace*{2.0cm}$F' \lthen x, \neg\neg x, H \Vdash_{G_3} M$ \hspace*{1.0cm} iff\\
\hspace*{2.0cm}$F' \liff x, \neg\neg x, H \Vdash_{G_3} M$\\

\noindent It is important to remark that $x$ does not occur in $H$ or in $F'$. For consistency,
the right implication clearly holds. We prove the left implication as follows.
We assume $F' \lthen x, \neg\neg x, H$ is consistent and complete. By the results obtained in \cite{tplp04,lopstr01}
$F' \lthen x, H$ is also consistent and complete. So there is an interpretation $I$ such that $I(H) = 2$. We
construct a new interpretation $I'$ that differs from $I$ only in that $I'(x) = I(F)$, so $I'$ models $F' \liff x, H$.
Therefore $F' \liff x, H$ is consistent. Since by hypothesis $F' \lthen x, \neg\neg x, H \Vdash_{G_3} x$ (because
$x \in M$) then clearly $F' \lthen x, H \Vdash_{G_3} x$. So $F' \liff x, \neg\neg x, H$ is consistent.\\

\noindent In respect to logical consequence, the right implication clearly holds. We will prove the left implication
for every atom $a \in M$. Again from the results obtained in \cite{tplp04,lopstr01} we can replace $F' \liff x, \neg\neg x, H$
by $F' \liff x, H$ in our proof, so we only need to prove\\

\noindent \hspace*{2.0cm}if $F' \liff x, H \vdash_{G_3} a$ then $F' \lthen x, H \vdash_{G_3} a$\\

\noindent We will prove its contrapositive, i.e.\\

\noindent \hspace*{2.0cm}if $\exists I$ $I((F' \liff x) \wedge H) > I(a)$ then 
$\exists I'$ $I'((F' \lthen x) \wedge H) > I'(a)$\\

\noindent If $a = x$ then we have
several cases. If $I((F' \lthen x) \wedge H) = 2$ and $I(a) = I(x) = 0$ then $I(F')$ must evaluate to 0, so 
with $I' := I$ the statement holds. The case when $I((F' \lthen x) \wedge H) = 1$ and $I(x) = 0$ is
analogous. For the case when $I((F' \lthen x) \wedge H) = 2$ and $I(x) = 1$, if $I(F') = 1$ then
we define $I' := I$ as in the previous cases. Now if $I(F') = 0$ we take $I'$ to be equal to $I$ except
for the fact that we define $I'(x) = 0$, which satisfies the desired condition. The proof for the case
when $x \neq a$ is analogous but simpler, and is therefore omitted. \mathproofbox

\bibliographystyle{acmtrans}
\bibliography{bibliography}

\end{document}